EEA Conference & Exhibition 2019,

25 – 27 June in Auckland

# A Reliability-Oriented Cost Optimisation Method for Capacity Planning of a Multi-Carrier Micro-Grid: A Case Study of Stewart Island, New Zealand

**Soheil Mohseni, Alan C. Brent, and Daniel Burmester**

Chair in Sustainable Energy Systems, School of Engineering and Computer Science, Victoria University of Wellington, New Zealand

Presenter:

Soheil Mohseni

**Abstract**

Nearly all types of energy systems – power systems, natural gas supply systems, fuel supply systems, and so forth – are going through a major transition from centralised, top-down structures to distributed, clean energy approaches in order to address the concerns regarding climate change, air quality, depletion of natural resources, and energy security, whilst also enabling the supply of energy to communities in line with the goals of sustainable development. Accordingly, the establishment of the concept of sustainable, decentralised, multi-carrier energy systems, together with the declining costs of renewable energy technologies, has proposed changes in the energy industry towards the development of integrated energy systems.

Notwithstanding the potential benefits, the optimal capacity planning of these systems with multiple energy carriers – electricity, heat, hydrogen, biogas – is exceedingly complex due to the concurrent goals and interrelated constraints that must be satisfied, as well as the heavily context-dependent nature of such schemes. This paper puts forward an innovative optimal capacity planning method for a cutting-edge, stand-alone multiple energy carrier micro-grid (MECM) serving the electricity, hot water, and transportation fuel demands of remote communities. The proposed MECM system is equipped with wind turbines, a hydrogen sub-system (including an electrolyser, a hydrogen reservoir, and a fuel cell), a hybrid super-capacitor/battery energy storage system, a hot water storage tank, a heat exchanger, an inline electric heater, a hydrogen refuelling station, and some power converters.

The objective of calculating the optimal sizes of the considered MECM through minimising its lifetime cost is subjected to the following constraints: (i) the fulfilment of some reliability indices for supplying the electrical, thermal, and transportation fuel load demands, (ii) the hourly balance between the generated and consumed energy on the MECM's network over its projected operational time frame, (iii) the provision of dynamic stability of the system, (iv) the equality of the initial and final states of energy reserves of the system, and (v) the compliance with the functional and technical characteristics of the employed components. A numerical case study for the optimal capacity planning of the suggested MECM configuration, to be realised on Stewart Island, New Zealand, is presented to evaluate the effectiveness of the proposed optimisation method.

# 1 Introduction

Under the Paris Agreement, the New Zealand government has committed to cutting greenhouse gas (GHG) emissions by 30% below 2005 levels, by 2030. Based on this commitment and the Productivity Commission's report, the Ministry for the Environment (MfE) is driving innovation in clean energy technologies and provisions [1], [2]. The introduction of renewable energy sources (RESs) and the advent of green transportation fleets are two major technological trends among the clean technologies in the energy sector that the MfE is pursuing to move New Zealand towards a sustainable future.

Micro-grids (MGs) are attracting considerable interest recently, due to their potential advantages in terms of facilitating the integration of RESs and green transportation technologies into the existing and upcoming energy systems. The so-called multiple energy carrier micro-grid (MECM) network is referred to as an interconnected energy system that provides a platform for the linking of different energy vectors from varied distributed energy resources (DERs) to meet a variety of energy needs in a region – electricity, heat, and transportation fuel [3]. The MECM model expands on the concept of "smart grids" with the aspiration of harnessing the interplay between different energy vectors to improve the resiliency, reliability, efficiency, and affordability of the renewable energy supply [4].

# 2 Motivation

The optimal equipment capacity planning of a MECM is pivotal for retaining the financial sustainability of the system, whilst ensuring reliable energy supply. This problem is a complex combinational optimisation problem due to its extremely large design space, the presence of several non-linear constraints involved in its formulation, and the lack of grid support against the fluctuations of the output powers from RESs. A novel method, based on artificial intelligence (AI) techniques, is thus needed to calculate the optimal capacity of the stand-alone MECMs' equipment [5]. Such a method will promote renewable energy systems successfully and effectively in remote areas by reducing the risk when investing in grid-independent renewable energy projects; thereby addressing New Zealand's climate change and sustainable development policies.

# 3 Mathematical Modelling of the System under Study

This section mathematically models the components integrated into the conceptualised stand-alone MECM system, whose schematic diagram and power flows are shown in Fig. 1, whilst providing insight into the operation of the network.

The output power of the wind turbine (WT) power plant at different wind speeds is modelled by the following equation [6]:

$$P_{WT}(t) = \begin{cases} 0 & ; V < V_{cin}, V > V_{cout} \\ P_{rated} \times ((V(t) - V_{cin})/(V_{rated} - V_{cin}))^3 & ; V_{cin} \leq V < V_{rated} \\ P_{rated} & ; V_{rated} \leq V \leq V_{cout} \end{cases} \quad (1)$$

where $V_{cin}$, $V_{cout}$, and $V_{rated}$ are the cut-in, cut-out, and nominal wind speeds, respectively; $V(t)$ is the wind speed at time step $t$; and $P_{rated}$ is the nominal power of the WTs. In this analysis, $V_{rated}$ and $P_{rated}$ are set to 15 m/s and 100 kW, respectively.

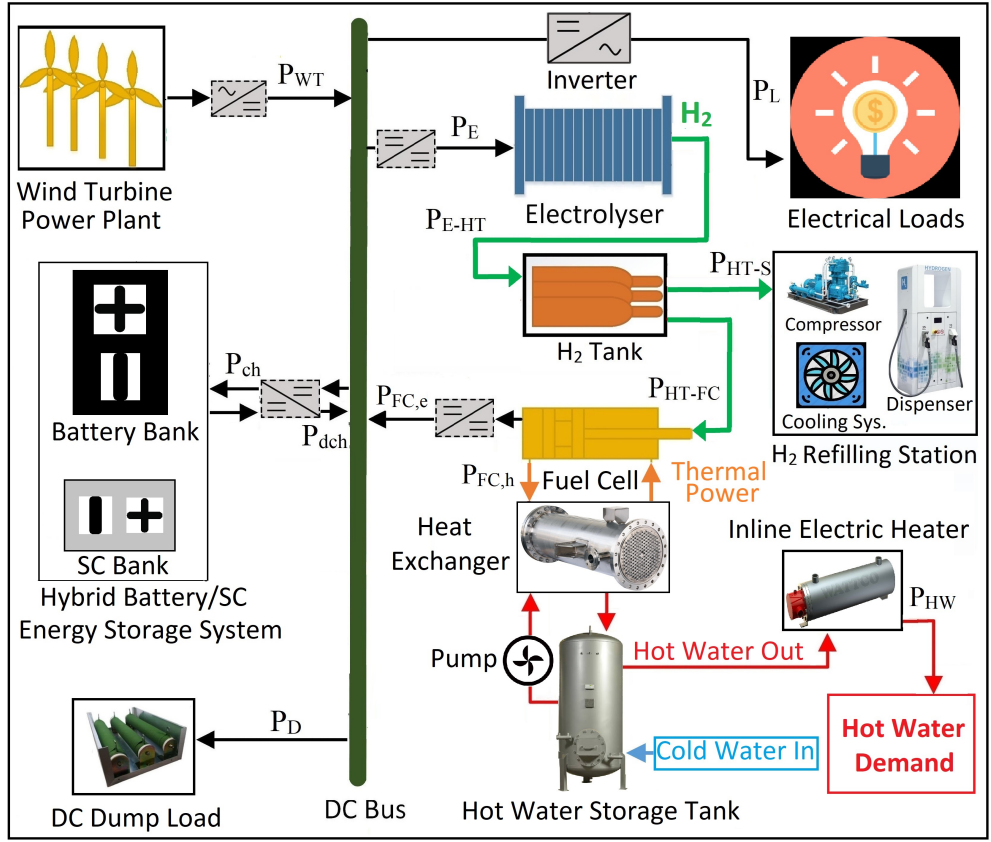

Fig. 1. Schematic diagram and power flow of the conceptualised MECM system.

The energy content of the hybrid super-capacitor (SC)/battery bank at each time step of the operation of the MG can be expressed by the following equation:

$$E_H(t) = E_H(t-1) + \left(P_{ch}(t) - \frac{P_{dch}(t)}{\eta_H}\right) \times \Delta t, \qquad (2)$$

where $\eta_H$ is the round-trip efficiency of the hybrid system (i.e. 92.5%), while $P_{ch}$ and $P_{dch}$ denote the charging and discharging powers of the system, respectively.[1]

The electrical output power of the polymer electrolyte membrane fuel cell (FC), which is driven electrically, and the waste heat is considered as its by-product (which is recovered using the heat exchanger), can be calculated by:

$$P_{FC,e} = \eta_{FC} \times P_{HT-FC}, \qquad (3)$$

where $\eta_{FC}$ is the FC's electrical efficiency, i.e. 50%. The amount of thermal power generated by the FC at a given operating power, $P_{FC,e}$, can be determined by [7]:

---
[1] The reader is referred to Fig. 1 for the description of the power flow parameters, where required.

$$P_{FC,h} = r_{FC}^h \times P_{FC,e}, \tag{4}$$

where $r_{FC}^h$ is the ratio of the fuel cell's thermal to electrical output power, i.e. 0.8. The water produced by the FC absorbs part of the $P_{FC,h}$, and the rest of it (here, 65% of $P_{FC,h}$) could be used for heat recovery purposes.

The amount of stored hydrogen (H$_2$) in the H$_2$ reservoir [kg] at time step $t$ can be obtained by:

$$m_{HT}(t) = (E_{HT}(t-1) + \left(P_{E-HT}(t) - \frac{(P_{HT-FC}(t) + P_{HT-S}(t))}{\eta_{tank}}\right) \times \Delta t)/HHV_{H_2}, \tag{5}$$

where $E_{HT}$ is the state of H$_2$ energy stored in the tank, $HHV_{H_2}$ is the higher heating value of the H$_2$, i.e. 39.7 kWh/kg, and $\eta_{tank}$ is the round-trip efficiency of the tank, i.e. 98%.

The amount of thermal energy (in the form of hot water) [kW] delivered from the hot water storage tank to the inline electric heater is calculated by [8]:

$$E_{HW}(t) = \dot{m}_{outlet} \times c_p \times \eta_{HW} \times (T_{out} - T_{in})/(3600), \tag{6}$$

where $\dot{m}_{outlet}$ is the mass flow rate of the hot water at the tank outlet [kg/h]; $c_p$ represents the specific heat capacity of water, i.e. 4.19 kJ/kg °C; $\eta_{HW}$ is the hot water tank's efficiency, i.e. 96%; with $T_{in}$ and $T_{out}$ respectively denoting the temperature of the water inflowing/outflowing to/from the hot water storage tank. In this study, $T_{in}$ is assumed to be constant at 12 °C.

An H$_2$ refuelling station is also integrated into the system, which serves the purpose of refilling the H$_2$ FC (HFC)-powered vessels and vehicles. It is comprised of a high-pressure compressor, a cooling system, and a dispenser.

Furthermore, the heat exchanger, the inline electric heater, the H$_2$ refuelling station, the polymer electrolyte membrane electrolyser, and the inverter are modelled by their efficiencies, which are set to 90, 97, 95, 60, and 90%, respectively, in the same way as outlined for the electrical output power of the FC in Eq. (3).

The temperature of the hot water demand is considered as 40 °C. In this regard, motivated by the idea proposed by Assaf and Shabani [8], an inline electric heater, which is powered (in order of preference) by the WT power plant or the FC, is used to heat up the water at the tank outlet to the desired temperature of the consumers (i.e. 40 °C), when necessary. Indeed, the inline electric heater acts as a back-up to the electrically-driven FC; as it is not reasonable to dump the excess electric output from the FC, which would be the case if the inline electric heater is not used as a resource to compensate for any lack of thermal power generation. A controller also ensures that the temperature of the water stored in the tank does not go beyond the acceptable limit of 65 °C.

The output power of the WT power plant, due to its weather-driven nature, varies temporally – seasonally, monthly, daily, and instantaneously. On the other hand, the variability in electric load demand can occur over a wide range of time scales, from seconds to months. Accordingly, three different back-up systems viz. the SC bank, the lithium-ion battery bank, and the stationary polymer electrolyte membrane fuel cell are considered to compensate for the volatility in load demand and/or the output power of the WT power plant. The rationale behind the use of these components lies in their different characteristics in terms of their energy and power densities [9]. Fuel cells/SCs are associated with high energy/power densities, but low power/energy densities; thus, they are best suited to overcome the mid-to-long term/instantaneous mismatches in renewable power supply and electricity demand. In addition, batteries bridge the gap between the SCs and FCs; they offer the prospect of compensating for the diurnal fluctuations in supply/demand owing to the intermediary level of both their energy and power densities.

The cycle-charging energy dispatch strategy is utilised to operate the conceptualised MECM. Accordingly, a low-pass energy filter initially decomposes the wind power-electricity demand mismatch signal into its low- and high-frequency components. The low-frequency component is then used to produce hydrogen using an electrolyser or to govern the operation of the FC, depending on the wind power excess or shortage; whereas, the high-frequency component passes through another low-pass energy filter with a lower cut-off frequency, as compared to the previous one. Subsequently, the low- and high- frequency components are adopted to charge/discharge (depending on the state of the power mismatch) the battery and SC banks (incorporated in the hybrid energy storage system), respectively. Any surplus power beyond the capacity of the electrolyser, $H_2$ reservoir, SC bank, and battery bank to deal with, is used in the inline electric heater to meet the hot water demands if there exists any as-yet-unsupplied thermal loads. In this situation, when the capacity of the heater is not adequate for meeting the thermal demand, the surplus power is dumped through a DC load consisting of a resistor bank – this increases the loss of thermal power supply probability reliability index. On the other hand, when the capacities of the battery bank, SC bank, or the electrolyser are not adequate for meeting wind power shortage (in electricity supply), a load-shedding scheme (which increases the loss of electric power supply probability) ensures the MG system's stability.

## 4 Reliability-Oriented Whole-Life Cost Minimisation

The proposed generic reliability-oriented life-cycle cost minimisation method seeks to improve the affordability, sustainability, and cost-efficiency of the stand-alone MGs supplying the electricity, transportation fuel, and hot water demands of remote communities. The proposed method is also capable of determining the optimum capacity of the MG's equipment by using the net present cost (NPC), the loss of power supply probability (LPSP), and the AI-based particle swarm optimisation (PSO) respectively for projecting the system's whole-life cost, measuring the system's reliability, and minimising the projected whole-life cost subject to some constraints.

The total NPC of the MG is the sum of all the NPCs of the MG's equipment. The NPC of each component can be calculated by [10]:

$$NPC = N \times \left( CC + RC \times K + \frac{O\&M}{CRF(d,R)} - SV \right), \qquad (7)$$

where $N$, $CC$, $RC$, $O\&M$, and $SV$ respectively denote the optimum capacity, capital cost, replacement cost, operation and maintenance cost, and salvage value of the component; $K$ is the single payment present worth; $CRF$ stands for the capital recovery factor; $d$ denotes the real interest rate per annum (i.e. 6%); and $R$ is the expected life span of the MG system, i.e. 20 years.

The total NPC of the system, $TNPC$, can also be annualised as follows:

$$TNPC_{ann} = CRF(d,R) \times TNPC. \qquad (8)$$

The MGs' levelised cost of energy [$/kWh] can then be calculated by dividing its annualised total NPC by the aggregate electric and thermal energy it serves to the customers over the planning horizon.

The minimisation of the MG's total NPC is carried out against relaxing the target reliability requirements to supply the electric, heat, and transportation fuel (i.e. $H_2$) energy demands. In this regard, three separate reliability indices based on the LPSP concept measure the consistency of the supply of electricity, hot water, and $H_2$ as a transportation fuel. The LPSP reliability indicator in meeting the electricity/hot water/transport fuel requirements can be determined by [11]:

$$LPSP = 100 \times \frac{\sum_{i=1}^{N} \text{hours}[P_{supp}(i) < P_{dem}(i)]}{N}, \quad (9)$$

where $P_{supp}$ could denote the supplied electric/thermal/$H_2$ power, $P_{dem}$ could denote the demand for electricity/hot water/transport fuel, and $N$ is the total number of hours considered in the time scale on which the MG is operated, i.e. 8760. For this project, the LPSP indices are set to 0%.

The non-convexity of the objective function, which is subject to several nonlinear constraints precludes the utilisation of exact mathematical optimisation methods to solve this problem. Accordingly, the PSO technique is utilised to minimise the total NPC of the MG, which was first proposed by Kennedy and Eberhart [12]. Compared to alternative AI-based optimisation techniques, the PSO is powerful because instead of using a 'survival of fittest' approach, the members of the PSO population interact and influence each other. In this analysis, in the PSO algorithm, the learning factors and inertia weight are assumed as 2 and 0.7, respectively; the population size is considered as 45; and the maximum number of iterations is assumed as 300.

In addition to the requirement of satisfying the specified reliability criteria, the optimisation problem is subject to some constraints on the: (1) equality of the initial and final states of energy stored in the battery bank, the SC bank, and the $H_2$ reservoir over an entire (8760-h) operational time frame; (2) the demand-supply balance for the electric power at each time step of operating the MG, as in Eq. (10); (3) retaining of the states of energy stored within the SC bank, battery bank, and the $H_2$ tank within their predefined permissible limits at each time slot of operating the MG; and (4) upper bounds of the design variables (capacity of the MG's equipment).

$$P_{WT}(t) + P_{ch}(t) + P_{FC,e}(t) = P_L(t) + P_{dch}(t) + P_E(t) + P_D(t) \quad \forall t \in \{1, 2, \dots, 8760\}. \quad (10)$$

Figure 2 shows an overview of the proposed method for the optimum investment planning of the conceptualised stand-alone MECM addressing various energy needs of a remote community.

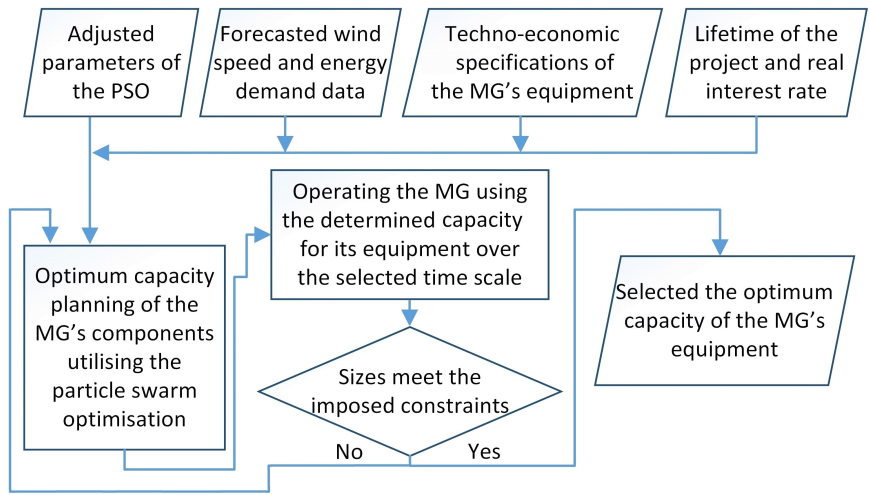

Fig. 2. Flowchart of the proposed optimum investment planning method for the devised MECM.

# 5 Input Data and Simulation Results

The proposed notional MECM network aims at providing clean, consistent electricity, transportation fuel, and hot water to a remote community of 400 people residing on Stewart Island, New Zealand. In a systematic study carried out on the assessment of the potentials for the utilisation of RESs on Stewart Island, Mason and McNeil [13] have suggested that the wind power generation is the unrivalled source of renewable power among a range of RESs (including solar, wind, hydro, and bio-diesel) due to the resource abundance and very small land and environmental permanent footprints, as compared to the other options available. Hence, a WT power plant is considered as the sole primary source of energy in the proposed stand-alone MECM architecture.

In order to forecast the hourly-basis, year-round output power profile of the WT power plant, real wind speed data were first collected from the New Zealand's National Institute of Water and Atmospheric Research (NIWA) CLiFlo database for Stewart Island over the years 2009 to 2018 [14], and then averaged at intervals of 1 h. The forecasted monthly mean 24-h profile for the wind speed [m/s] is shown in Fig. 3.

The hourly-basis, year-round electrical load power profile is forecasted based on the New Zealand GREEN grid household electricity demand study, which incorporates the space heating energy demand [15]. The hourly-basis, year-round thermal load power (i.e. hot water demand) profile is forecasted according to the studies conducted in [16], assuming that each person uses 44 litres of hot water per day. The monthly mean 24-h profiles for the forecasted electric and thermal load powers [kW] are shown in Figs. 4 and 5, respectively.

Also, the typical daily profile for the $H_2$ load [kg $H_2$/h] imposed on the conceptualised MECM model, with the goal of decarbonising the transportation sector, is shown in Fig. 6. The following assumptions were made in deriving this $H_2$ load pattern: (1) Five HFC-powered ferries, five HFC-powered trucks, and five HFC-powered tractors, which, respectively, can hold 31.7, 8.2, and 32.9 kg of $H_2$ in their carbon composite tanks have to be integrated into the system. The marine vessels serve the purpose of transporting the passengers between Stewart Island and the Bluff, while the trucks and tractors effectively contribute towards achieving the objectives of agricultural sustainability; (2) a fleet of thirty 8.5-kW light-duty HFC-powered vehicles utilise the $H_2$ station to refill their 1.5-kg $H_2$ tanks; (3) a valley-filling energy management scheme that refuels the vessels and heavy-duty vehicles (i.e. trucks and tractors) in the early morning hours (by uniformly distributing their $H_2$ loads over the hours 1 to 6 am) is adopted, while the light-duty vehicles utilise the station during day-time hours (from 9 am to 8 pm) following an appropriate Normal distribution; and (4) the $H_2$ tanks of the ferries, light-duty vehicles trucks, tractors, and trucks need to be refuelled from 5 to 100% of their capacities at every 2, 3, 4, and 5 days, respectively. Moreover, the techno-economic specifications of the conceptualised MG's equipment are presented in Table 1 [4], [7], [8], [16]. State-of-the-art components are used in the MG's structure, whose costs are reported in U.S. dollars in Table 1.

The modelling, optimisation, and analysis of the conceptualised MECM system were carried out using the MATLAB® software. The optimum combination of the capacity of the MG's equipment obtained by solving the formulated problem using the PSO algorithm subject to the imposed constraints is presented in Table 2. The minimised total NPC of the MG system is found as US$7,961,243, which is broken down in terms of its constituent cost components and graphically depicted in the form of a radar chart in Fig. 7. Although the numbers shown on the

graph represent the real NPCs associated with the optimum capacities of the MG's equipment, the chart is plotted on the logarithmic scale for the sake of better visualisation.

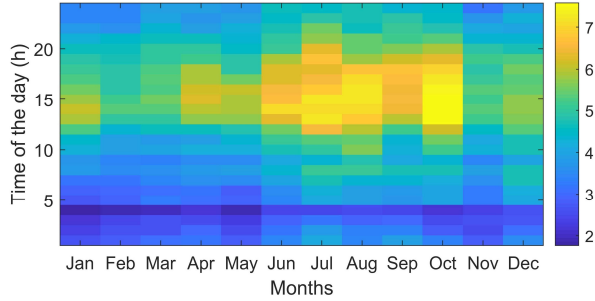

Fig. 3. Monthly mean daily wind speed [m/s].

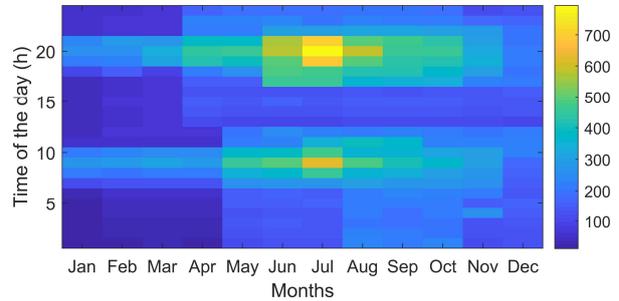

Fig. 4. Monthly mean 24-h electric load [kW].

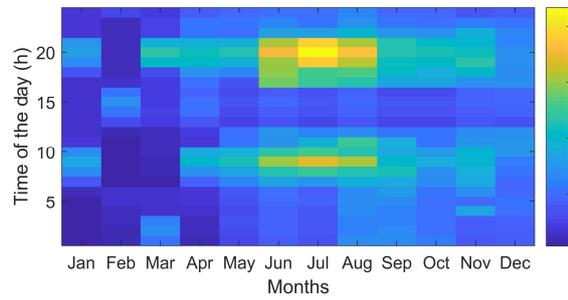

Fig. 5. Monthly mean 24-h heat load [kW].

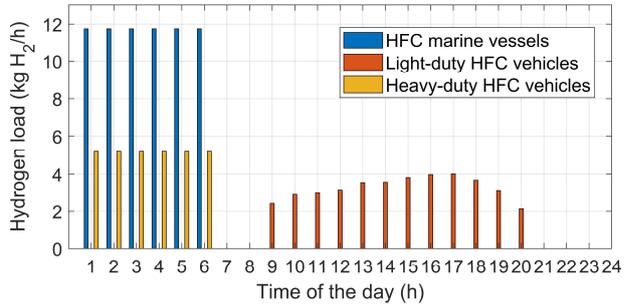

Fig. 6. Typical daily $H_2$ load profile [kg $H_2$/h].

Table 1. Techno-economic specifications of the MG's equipment.

| Component | Rated capacity | $CC$[a] | $RC$[a] | $O\&M$[a] | Efficiency[b] [%] | Lifetime [years] |
|---|---|---|---|---|---|---|
| WT | 100 kW | $120k/unit | $100k/unit | $4.6k/unit/year | N/A[c] | 20 |
| SC modules | 3.23 Wh | $32/module | $32/module | $0.5/module/year | 95 | 10 |
| Battery packs | 7.5 kWh | $630/pack | $600/pack | 20/pack/year | 90 | 12 |
| Electrolyser | generic | $1k/kW | $1k/kW | $20/kW/year | 60 | 15 |
| $H_2$ reservoir | generic | $470/kg | $470/kg | $9/kg/year | 98 | 20 |
| Fuel cell | generic | $1.1k/kW | $0.9k/kW | $28/kW/year | 50[d] | 5 |
| Heat exchanger | generic | $100/kW | $90/kW | $2/kW/year | 90 | 15 |
| Hot water tank[e] | generic | $0.5/L | $0.3/L | $0/L/year | 96 | 15 |
| Inline electric heater | generic | $1k/kW | $1k/kW | $8/kW/year | 97 | 15 |
| $H_2$ refilling station | generic | $6k/kg $H_2$/h | $5k/kg $H_2$/h | $180/kg $H_2$/h/year | 95 | 20 |
| Electric loads' inverter | generic | $350/kW | $300/kW | $7/kW/year | 90 | 15 |

[a] The costs include the costs associated with the converters shown inside the dashed lines in Fig. 1.
[b] The equipment efficiency is reported considering the efficiencies associated with the converters shown inside the dashed lines in Fig. 1.
[c] The WT plant is modelled using its power curve, which governs the relationship between its output power and the hub height wind speed.
[d] The value represents the fuel cell's electric efficiency.
[e] The hot water tank's specifications include the specifications associated with the water pump shown in Fig. 1.

Table 2. Optimised capacity of the MG's equipment using the PSO technique.

| WTs [no.] | SC modules [no.] | Battery packs [no.] | Electrolyser [kW] | $H_2$ reservoir [kg] | Fuel cell [kW] | Heat exchanger [kW] | Hot water tank [L] | Inline heater [kW] | $H_2$ station [kg $H_2$/h] | Inverter [kW] |
|---|---|---|---|---|---|---|---|---|---|---|
| 31 | 8,376 | 18 | 964 | 619 | 261 | 213 | 283,301 | 97 | 17.2 | 741 |

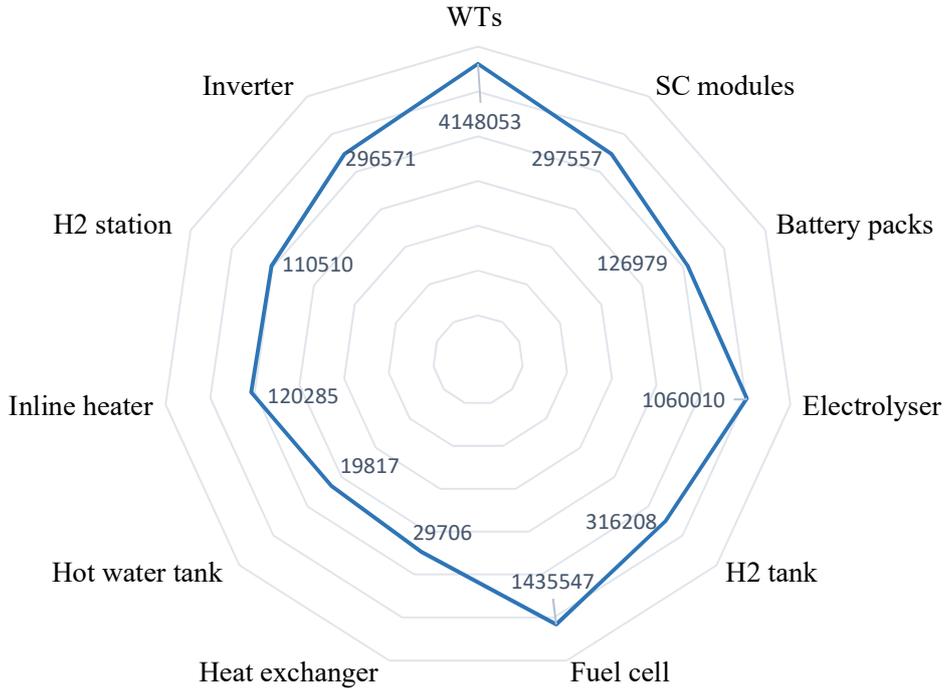

Fig. 7. Breakdown of the optimised total NPC of the conceptualised MECM [US$].

In order to evaluate the financial viability of realising the proposed MECM for Stewart Island, the levelised cost of energy of the MECM model is estimated by dividing the projected total NPC by the total amount of energy it feeds to the electric, heat, and $H_2$ loads during its life span of 20 years, which is found as $0.27/kWh, in New Zealand currency. The levelised cost of energy is then split into the levelised costs of electricity, heat, and hydrogen, exploiting the fact that the NPCs associated with the components that take part in producing only one form of energy for end-use consumption (e.g. $H_2$ station, inline heater, or inverter) should not be considered in formulating the levelised costs of the other energy forms. For the components that take part in producing more than one energy form (i.e. the fuel cell), the NPCs are split according to their contributions to serving each type of end-use loads. Accordingly, the levelised costs of electricity, hot water (for direct use), and hydrogen production using the conceptualised 100% renewable MECM for Stewart Island are determined as NZ$0.20/kWh, NZ$0.0091/L, and NZ$6.97/kg $H_2$, respectively.

Presently, the electricity on Stewart Island costs about NZ$0.26/kWh, on average. The existing power system is entirely based on diesel generators on a centralised basis. Furthermore, the most recent studies on the renewable $H_2$ production in New Zealand have reported the levelised costs of NZ$14/kg $H_2$ and NZ$8.91/kg $H_2$ respectively for the small- and large-scale $H_2$ production plans [17], [18]. Moreover, in general, depending on the availability of RESs, system scale, and technologies utilised to heat the water renewably, a litre of hot water is expected to cost somewhere between 0.0077 and NZ$0.028 [19].

Based on the above premises, the proposed MECM system, if realised, would impose lower electricity charges on the customers than the existing non-renewable power system on the island. It would also produce hydrogen at rates well below the state-of-the-art architectures, which could facilitate the transition towards a low-carbon transportation system. Moreover, it would be able to satisfy the residential needs for hot water at a rate highly competitive with those of the cutting-edge renewable technologies and systems developed for water heating. Thus, it could be concluded that the proposed MG system introduces a cost-effective plan to realise the targets of decarbonising the island's energy sector; whilst also building energy resilience, achieving energy self-sufficiency, enhancing energy security, and meeting the energy supply reliability requirements.

In order to further validate the long-term economic viability of implementing the suggested MG project on the island, a thorough cost-benefit analysis is carried by the following three financial metrics: (1) the discounted payback period (DPP), (2) the profitability index (PI), and (3) the internal rate of return (IRR). Table 3 presents the obtained values for these economic indicators.

Table 3. Economic sustainability evaluation of the MG project under study.

| DPP (year) | PI (%) | IRR (%) |
|---|---|---|
| 8.79 | 2.45 | 13.68 |

The obtained values for the considered financial viability measures imply that the proposed renewable energy project is not only financially sustainable, but also could be identified as a profit-making, low-risk opportunity for investment, which creates a steady revenue stream and makes a decent return on capital without the need for subsidies. It is conceivable that the MG's ownership structure itself could affect the financial assessments conducted. For example, classifying the consumers as shareholders would require the system designer to include the subsidised costs associated with the hydrogen vehicles/vessels into the model by using appropriate multi-stakeholder business models integrating the government, the system operator, investors, and consumers.

# 6   Conclusions

This paper highlights the socio-economic benefits of developing a sustainable, carbon-neutral, stand-alone MG system (where $H_2$ is utilised as an energy vector), satisfying nearly all the energy needs (i.e. electricity, space heating, hot water, and transportation fuel) of a remote community residing on Stewart Island, New Zealand. The conceptualised MECM system is equipped with a hybrid energy storage system consisting of an SC bank, a battery bank, and an $H_2$ reservoir to back up the system respectively in the face of instantaneous, intra-hour, and intra-day variability of the system's status, introduced by intermittent loads and or/wind power generation. A novel operational strategy is also devised to cost-efficiently integrate the fleets of HFC-powered light-duty vehicles, heavy-duty vehicles, and ferries, into the proposed 100% renewably powered MECM model, which thereby holds the potential of paving the way towards realising a green transportation system for the island. The proposed MECM system could be easily adapted for use across New Zealand, either by incorporating other renewable energy technologies (e.g. PV panels, micro-hydro power plants, biomass power plants, etc.), following an analysis of the significant potential RESs at the case study site and/or adding a grid interface to the MG layout to connect it to the National Grid. Hence, it can provide a credible path forward in advancing New Zealand's transition to a sustainable, low-carbon energy economy.